\pdfoutput=1
\documentclass[english,runningheads,a4paper]{llncs}
\usepackage[english]{babel}
\addto\extrasenglish{\languageshorthands{english}\useshorthands{"}}

\usepackage{regexpatch}
\makeatletter
\edef\switcht@albion{%
  \relax\unexpanded\expandafter{\switcht@albion}%
}
\xpatchcmd*{\switcht@albion}{ \def}{\def}{}{}
\xpatchcmd{\switcht@albion}{\relax}{}{}{}
\edef\switcht@deutsch{%
  \relax\unexpanded\expandafter{\switcht@deutsch}%
}
\xpatchcmd*{\switcht@deutsch}{ \def}{\def}{}{}
\xpatchcmd{\switcht@deutsch}{\relax}{}{}{}
\edef\switcht@francais{%
  \relax\unexpanded\expandafter{\switcht@francais}%
}
\xpatchcmd*{\switcht@francais}{ \def}{\def}{}{}
\xpatchcmd{\switcht@francais}{\relax}{}{}{}
\makeatother

\usepackage{ifluatex}
\ifluatex
  \usepackage{fontspec}
  \usepackage[english]{selnolig}
\fi

\iftrue 
  \ifluatex
  \else
    \usepackage[%
      rm={oldstyle=false,proportional=true},%
      sf={oldstyle=false,proportional=true},%
      tt={oldstyle=false,proportional=true,variable=false},%
      qt=false%
    ]{cfr-lm}
  \fi
\else
  \ifluatex
  \else
    \usepackage{newtxtext}
    \usepackage{newtxmath}
    \usepackage[zerostyle=b,scaled=.9]{newtxtt}
  \fi
\fi

\ifluatex
\else
  \usepackage[T1]{fontenc}
  \usepackage[utf8]{inputenc} 
\fi

\usepackage{graphicx}
\usepackage{upquote}
\usepackage{booktabs}
\usepackage{paralist}
\usepackage{csquotes}
\usepackage{textcmds}

\RequirePackage[%
  babel,%
  final,%
  expansion=alltext,%
  protrusion=alltext-nott]{microtype}%
\DisableLigatures{encoding = T1, family = tt* }

\usepackage{url}
\makeatletter
\g@addto@macro{\UrlBreaks}{\UrlOrds}
\makeatother
\usepackage{xcolor}

\usepackage{listings}
\renewcommand{\lstlistingname}{List.}
\usepackage{chngcntr}
\AtBeginDocument{\counterwithout{lstlisting}{section}}

\usepackage{pdfcomment}
\usepackage[%
  square,        
  comma,         
  numbers,       
  sort&compress, 
]{natbib}

\ifluatex
\else
  \SetExpansion
  [ context = sloppy,
    stretch = 30,
    shrink = 60,
    step = 5 ]
  { encoding = {OT1,T1,TS1} }
  { }
\fi

\usepackage{stfloats}
\fnbelowfloat
\usepackage{hyperref}
\hypersetup{hidelinks,
	colorlinks=true,
	allcolors=black,
	pdfstartview=Fit,
	breaklinks=true}
\usepackage[all]{hypcap}

\usepackage[group-four-digits,per-mode=fraction]{siunitx}

\usepackage[capitalise,nameinlink]{cleveref}
\usepackage{iflang}
\IfLanguageName{ngerman}{
  \crefname{table}{Tab.}{Tab.}
  \Crefname{table}{Tabelle}{Tabellen}
  \crefname{figure}{\figurename}{\figurename}
  \Crefname{figure}{Abbildungen}{Abbildungen}
  \crefname{equation}{Gleichung}{Gleichungen}
  \Crefname{equation}{Gleichung}{Gleichungen}
  \crefname{listing}{\lstlistingname}{\lstlistingname}
  \Crefname{listing}{Listing}{Listings}
  \crefname{section}{Abschnitt}{Abschnitte}
  \Crefname{section}{Abschnitt}{Abschnitte}
  \crefname{paragraph}{Abschnitt}{Abschnitte}
  \Crefname{paragraph}{Abschnitt}{Abschnitte}
  \crefname{subparagraph}{Abschnitt}{Abschnitte}
  \Crefname{subparagraph}{Abschnitt}{Abschnitte}
}{
  \crefname{section}{Sect.}{Sect.}
  \Crefname{section}{Section}{Sections}
  \crefname{listing}{\lstlistingname}{\lstlistingname}
  \Crefname{listing}{Listing}{Listings}
}

\usepackage{xspace}

\DeclareFontFamily{U}{MnSymbolC}{}
\DeclareSymbolFont{MnSyC}{U}{MnSymbolC}{m}{n}
\DeclareFontShape{U}{MnSymbolC}{m}{n}{
  <-6>    MnSymbolC5
  <6-7>   MnSymbolC6
  <7-8>   MnSymbolC7
  <8-9>   MnSymbolC8
  <9-10>  MnSymbolC9
  <10-12> MnSymbolC10
  <12->   MnSymbolC12%
}{}
\DeclareMathSymbol{\powerset}{\mathord}{MnSyC}{180}

\ifluatex
\else
  \input glyphtounicode
\fi
\usepackage[math]{blindtext}
\usepackage{mwe}

\usepackage{graphicx}

\usepackage{perpage} 
\MakePerPage{footnote} 

\begin{document}
\title{JSSignature: Eliminating Third-Party-Hosted JavaScript Infection Threats Using Digital Signatures}
\titlerunning{JSSignature: Eliminating Third-Party-Hosted JavaScript Infection Threats...}

%
%
\author{
Kousha Nakhaei\inst{1} 
\and 
Ebrahim Ansari\inst{2,3}
\and
Fateme Ansari\inst{3}
}
\authorrunning{K. Nakhaei, E. Ansari and F. Ansari}
%
\institute{
MouseStats Analytics Inc., Vancouver BC V5A1M5, Canada,\
\email{k@mousestats.com}
https://kousha.ca/
\and
Charles University, Faculty of Mathematics and Physics, \\Institute of Formal and Applied Linguistics
\and
Department of Computer Science and Information Technology,\\
Institute for Advanced Studies in Basic Sciences (IASBS), Zanjan, Iran\\
\email{\{ansari,f.ansari\}@iasbs.ac.ir}
\\https://iasbs.ac.ir/~ansari
}

\maketitle              
\begin{abstract}
Today, third-party JavaScript resources are indispensable part of the web platform.  More than 88\% of world's top websites include at least one JavaScript resource from a remote host.  However, there is a great security risk behind using a third-party JavaScript resource, if an attacker can infect one of these remote JavaScript resources all websites those have included the script would be at risk.

In this paper, we present JSSignature, an entirely at the client-side pure JavaScript framework in order to validate third-party JavaScript resources using digital signature. Therefore, all included JavaScript resources are checked against the integrity, authentication and non-repudiation risks before the execution. In contrary to existing methods, JSSignature protects web pages regardless of third-party resource infection nature while it does not set any restrictions on trusted JavaScript providers. This approach has an acceptable one-time performance overhead and is an easily deployable add-in. We have validated the proposed solution by applying tests on an implemented version\footnote{The source-code, resources and the working demo are available at http://iasbs.ac.ir/\~{}ansari/jssignature/demo.html}.

\end{abstract}

\section{Introduction}
In recent years, web platform has become the primary platform on the Internet, and sequentially one of the most important software ecosystems. There is a significant interest in making rich client-side web applications, thanks to the new browser features and the overall demand. There are many good reasons to make rich client-side dynamic web applications. For example, HTML5 has introduced lots of useful features such as Web Workers \cite{w3c:workers} and Canvas \cite{w3c:canvas} which has made HTML as an appealing client-side platform for developers. 

To provide rich client-side web pages, JavaScript as a dynamic client-side programming language is inseparable part of all modern web applications. According to \cite{jsuage}, in 2014 87.8\% of all websites are using JavaScript as a client-side language.

Each single web page can include several external JavaScript resources. A developer may include third-party JavaScript libraries to unlock some features and advantages. For instance, using external JavaScript libraries for visitor statistics and analysis purposes are popular (e.g., Google Analytics, Woopra). Developers may use JavaScript libraries to improve user experience or make client-side programming easier (e.g., jQuery, MooTools, AngularJs). Recently, social media services have provided third-party JavaScript libraries to integrate social features within websites (e.g. Facebook Like and sharing, Google+1). 

There are two methods to include a JavaScript resource: (1) developer can make a clone of JavaScript library and serve it from his server, (2) and the second method is to include it from the vendor's server as a third-party-hosted JavaScript; The webpage visitor web browser will download the script from the third-party server.

In most cases developers prefer the latter to achieve its advantages: (1) CDNs and caching are major enthusiasm to use third-party hosted JavaScript library. Using a shared JavaScript host for popular libraries will decrease the page loading time in all websites those include a shared resource. The web browser will use a cached version of a JavaScript library instead of downloading it again for each website. (2) On the other hand, the use of a self-hosted clone of a library would increase maintenance cost since JavaScript libraries may be updated regularly. By each update, all websites need to keep their libraries updated manually by uploading the latest version of the libraries to their resource provider. (3) Content request distribution is another reason to use remotely hosted resources. A website owner may decide to host its website static resources in different networks or by a CDN provider to decrease the local web server stress, cost or load time. (4) After all, using a self-hosted JavaScript resource is impossible when the used third-party service is based on the frequent JavaScript resource updates (e.g. Google Tag Manager).

For each visitor, the eligible website may include some JavaScript codes from different vendors. All these JavaScript codes will be downloaded and executed in client's browsers with the same permission of the eligible web page, in a same execution space. Consequently, the third-party JavaScript codes has a full access to the web page DOM and other resources in the client-side.

Regarding the prevalence of third-party JavaScript libraries usage, based on a recent research, 88.45\% of world's top 10,000 websites are using at least one remote JavaScript library \cite{youarewhatyouinclude}. The tremendous growth in web-based applications has made web platform a major target for hackers. 

As demonstrated in Figure 1, if an attacker could insert an infected code into a third-party JavaScript resource, then all eligible websites that include the infected resource will be faced with serious security risks.

The situation will become worse by spread of particular third-party JavaScript providers those are used by thousands of world top websites. According to a report in 2014, there are about 11 million websites those are just using Google Hosted Libraries as a CDN JavaScript provider \cite{cdnstats}. It has become a serious security breakpoint for the whole web ecosystem. A possible attack on the hosted JavaScript on this CDN, will endanger millions of people information on all websites those are using these libraries. As an example, the hypothetical attacker is able to hijack the visitor session by using a few lines of code injected to a third-party library, to get all credit card details on payment pages or track login details of visitors through cookies or DOM  manipulation in all websites that are using the vulnerable hosted JavaScript resource.

Moreover, there are also more abuse cases which will make the third-party library as an appealing target for attackers. They can exploit recent browser vulnerabilities or distribute malware by \textit{drive-by-downloads} attacks through JavaScript codes in millions of websites and clients just by attacking to a single host. The mentioned reasons will make all third-party JavaScript providers as an appealing target for attackers.

The only inhibiting factor related to this issue which is implemented in all modern web browsers is Same-Origin Policy (\textit{SOP}). \textit{SOP}  permits each page's script to access all resources on same origin \cite{w3c:sop}. In other words, it will eliminate any DOM request from the script in website A to website B. By including JavaScript from Website B in website A, it will be treated as a script that belongs to website A, therefore it has a full permission to all elements and resources in website A, and \textit{SOP} would not prevent any risk. Therefore, it is not designed to solve the included third-party JavaScript security concern.

\begin{figure}[ht!]
\centering
\includegraphics[width=70mm]{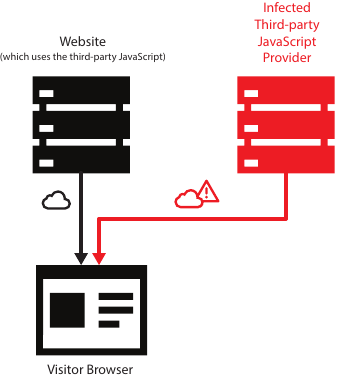}
\caption{Security concern. Infection of a third-party JavaScript will endanger all web applications those are using the third-party library, consequently all web application users are in danger. \label{overflow}}
\end{figure}

To clarify, by injecting the infected code into a benign third-party JavaScript resource, the attacker is able to simply steal all the private information. He can transfer data using indirect communication methods such as creating an image or script element on the fly with custom URL containing private information as a URL query. \textit{SOP} has no suggestion to protect the webpage from a third-party JavaScript inclusion infection.

Several studies have been presented to tackle the before mentioned security threat of third-party JavaScript resources. A recent study \cite{youarewhatyouinclude} has discussed two countermeasures to be protected from third-party JavaScript threats: (1) Proper sandboxing of third-party JavaScript, (2) Using local copy and avoid using third-party hosted scripts. All solutions are around sandboxing or limiting execution of third-party JavaScript.  They are working in different ways, (1) JSand \cite{jssand}, WebJail \cite{webjail}, ConScript \cite{conscript}, AdJail \cite{adjail} and AdSentry \cite{adsentry} provide a refined access control for JavaScript to define access policies for each library, (2) Caja \cite{caja}, ADsafe \cite{adsafe} and GateKeeper \cite{gatekeeper} are dedicated to make a safe sub-language by rewriting the script into a safe subset of codes and scripts or compare the script with a safe subset \cite{adsafe}.

However, the proposed solutions have severe drawbacks about solving the discussed security issues, (1) abusing trusted policies is the main concern. An attacker can utilize current whitelisted operations to do the attack, (2) also defining policies for each library need to know all eligible usages. Detecting all the functionalities of a large JavaScript library even if be possible, is a time consuming task. Moreover, in real world applications, trusted vendor may change his script time by time to add or remove features. They may completely rewrite the script to improve it. So, the policies may change over time. Consequently, keeping policies up-to-date will vanish the major advantage of third-party JavaScript regarding the decreased maintenance cost, (3) all suggested solutions are trying to implement a method similar to a firewall between web page and third-party JavaScript. At the end, they are restricting the benign third-party JavaScript functionality all the time due to the risk of possible attack to the vendor. (4) Moreover, some proposed methods require browser modification \cite{webjail, conscript, adsentry} which is not desirable in this case. All the web application users should install a modified browser to be protected against this kind of attacks. (5) Majority of the suggested workarounds have a significant process overhead because of complicated preprocessing or runtime monitoring which is not suitable for heavy JavaScript libraries, (6) After all, even if we solve the above mentioned issues, possible vulnerability in these proposed solutions will remain a concern \cite{languageisolation}. There is no guarantee these methods are invulnerable against all malicious JavaScript codes as their protective method is dependent on the included JavaScript codes functionality.

This paper presents JSSignature, a novel method to bring digital signatures to third-party JavaScript to eliminate security concern around third-party JavaScript inclusion. This method is inspired by a recent study on the prevalence of third-party JavaScript usage and current security obstacles \cite{youarewhatyouinclude}.

This paper makes the following contributions in the field:

\begin{enumerate}

\item It introduces JSSignature as a simple and straight-forward method that: (1) works without defining policies, consequently white-listed rules abuse is no longer disputable, 
(2) can work with all libraries and scripts independent of future updates without time consuming settings, 
(3) completely protect third-party JavaScript users from unauthorized script modifications without restricting the third-party JavaScript,
(4) works without browser modification or any modification related to the web page visitors,
(5) has just an optimum first load overhead and no impact on performance after initiation, 
(6) and thanks to the mathematical scheme, it works as an invulnerable solution independent of third-party script infection nature.
\item  We experimentally demonstrate the performance efficiency of our suggested method.
\item  Evidences will be provided to prove statements by implementing a prototype.

\end{enumerate}

The rest of this paper is structured as follows. We mention more details about the problem background in Section 2. Later, Section 3 is dedicated to present JSSignature architecture and system design. In Section 4, we will evaluate performance, security and compatibility of JSSignature. At the end, Section 5 and 6 in order are assigned to review related works and conclusion.

\section{PROBLEM STATEMENT}
While using third-party-hosted JavaScript resources has become an indispensable part of millions of websites, it can be considered as a serious threat for the whole web platform with the currently available measures.

A benign third-party JavaScript provider can be infected by a malicious code by just injecting a few lines of JavaScript code into a single provided resource. Consequently, the malicious code will be indirectly injected to all websites those have included the previously trusted JavaScript resource. The malicious code will be promptly executed for all the websites visitors. The whole procedure would happen in real-time for all websites and their users.

\subsection{Third-party JavaScript Inclusion}
The current security methods in modern web browsers to overcome the third-party JavaScript security issues are based on a binary strict choice. It means there are two options for using a third-party script, (1) full isolation by using iframe sandboxing, or (2) full access to web page by inclusion of the JavaScript using script tag.

The former completely isolates the script from accessing to the parent according to the \textit{SOP} or sandbox iframe HTML attributes. Full isolation disregards the reasons behind using a third-party JavaScript. The vast majority of JavaScript libraries is implemented to manipulate DOM or other resources of the web page. Obviously, full isolation of the script is not possible in this case.
 
The latter option includes third-party JavaScript resource using $<$script$>$ tag, and it remains as the only choice to utilize a third-party JavaScript resource. This paper takes into account including script using $<$script$>$ tag. This tag grants a full access of third-party JavaScript to the web page context and execution environment.

\subsection{Possible Attacks}
By super strong incentives for the attackers, the third-party JavaScript vendor is an extremely valuable target.
The attacker is able to have access to all data and functions in a web page through a JavaScript inclusion. Some accessible data in web pages are: cookies, DOM (e.g., Text and password input values), JavaScript variables, Web Storage, IndexedDB, etc. To clarify, the attacker would have access to DOM space, and read all the entered authentication credentials directly from the login fields. As another simple attack, they can hijack user session by bypassing web browser Cookie data.

Second major motivating factor is exploiting \textit{zero-day} vulnerabilities using malicious scripts. The benign third-party script provider can be used as an attack vector. Due to the fact that each third-party JavaScript may be used by many web applications, and each application can have millions of visitors, the target value as an attack vector is beyond estimation.

A trusted thirty-party JavaScript library can be infected in different ways. An attacker may (1) gain access to vendor's hosting and inject malicious script into the benign JavaScript library, (2) serve his own infected JavaScript instead of benign script using network attack methods (e.g., DNS hijacking, Domain hijacking, Domain sniping, Man-in-the-middle attack).

\subsection{Requirements}
Regarding the wide-spread usage of third-party JavaScript inclusions and considering serious security issues, the solution should satisfy the following requirements as essential principles:

\begin{enumerate}[a.]

\item \textbf{Invulnerable}. Completely eliminate any unauthorized third-party JavaScript modification. Protecting third-party script end-users from any kind of attack on the remote script contents (e.g., abusing white-listed policies, browser vulnerability exploits, the network routing attacks) is a must. 
\item \textbf{Keep third-party JavaScript advantages}. For instance, updating the third-party JavaScript resource should not force all users to change their settings (e.g., updating the hash checksum, changing the white-listed policies, uploading a new version of library). The solution should not set any restriction for a trusted third-party JavaScript provider, either in the current version or future developments. On the other hand, a third-party JavaScript must be loaded from provider's host to achieve CDN advantages and caching.
\item \textbf{Efficient performance}. Any performance penalty is not acceptable while executing the third-party JavaScript code. The performance penalty on JavaScript resource initialization must be minimum and unnoticeable for the end-user.
\item \textbf{Backward-compatible}. Utilizing a security method must be supported in all standard browsers without any modification. It should be implementable in the contemporary web platform without changing infrastructures. It should be compatible with current JavaScript libraries, without requiring to rewrite or modify any code.
\end{enumerate}

\section{PROPOSED SECURITY SCHEME}
In this paper, we exploit digital signatures' properties to provide a solution for third-party-hosted JavaScript resources threats. JSSignature brings three properties as \textit{integrity}, \textit{authentication} and \textit{non-repudiation} to the external JavaScript inclusions by using \textbf{digital signature} scheme. The website owner is able to include a trusted or manually reviewed third-party JavaScript without fear of possible attack against JavaScript provider.

\subsection{JSSignature Architectural Overview}
\begin{figure}[ht!]
\centering
\includegraphics[width=80mm]{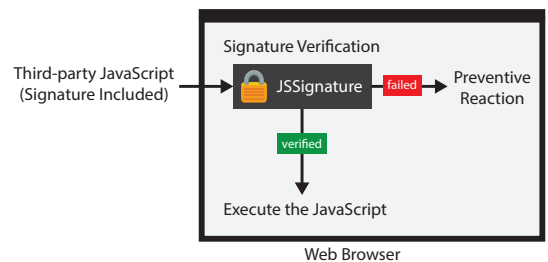}
\caption{JSSignature diagram. JSSignature verifies the Third-party JavaScript resource in the client-side before execution using digital signature scheme.\label{overflow}}
\end{figure}
Figure 2 illustrates the scheme of the proposed security method for the use case of external JavaScript inclusion. The web developer includes a self-hosted JSSignature library in the web page. Instead of using the script tag to include external script, the duty of loading third-party JavaScript resources is assigned to JSSignature. Thus, JSSignature acts similar to a client-side proxy to load the third-party JavaScript. It will fetch external script as a plain-text by using JavaScript functions in the client-side and after validating the content, it decides if the downloaded script should be executed or not.

\subsection{JSSignature Workflow}
JSSignature generates the signature to be used by a third-party JavaScript provider for signing procedure in the following steps: 
(1) make the hash of JavaScript content, 
(2) and create the signature by the asymmetric encryption of the calculated hash using the private key. The signing procedure should be done on a separate isolated computer, and not a public server. Therefore, the private key will be kept securely regardless of servers vulnerability.

JSSignature's workflow in a visitor's browser are as the following:
(1) fetch the third-party JavaScript resource as a plain-text, 
(2) find the signature on the first line of the downloaded script, 
(3) decrypt the signature using the public key which is included in the parent webpage, 
(4) hash the rest of the script using a cryptographic hash function, 
(5) compare calculated hash with the decrypted signature and decide to execute or go for a preventive reaction (e.g., block script, run locally hosted copy) based on this comparison.

For legacy support in case the JavaScript provider has not attached a signature to their script, JSSignature works a bit different: 
(1) fetch the script, 
(2) compute hash, 
(3) compare computed hash with the hash that the website owner has specified to decide whether the script is the validated version or not.

There are some other optional features JSSignature can benefit from:
\begin{enumerate}[a.]

\item \textbf{Certificate Authorities}. 
There is no technical barriers to transform the signing procedure and validation check to a Certificate Authority (\textit{CA}) based algorithm. It introduces an extra level of security as there is no need to share the public key securely and separately. Public key can be included in the signed resource. However, the signing procedure would require \textit{CA} certificate issuance cost for the third-party JavaScript provider. 

\item  \textbf{Multiple Signatures}. Having a JavaScript resource to be signed by different identities is possible and introduces more extra level of security. For example, a JavaScript resource can be signed twice by (1) the developer company, (2) and \textit{CDN} host provider. In this case, the resource can not compromised even by disclosing a single identity's private key. 

\end{enumerate}
In this paper, we describe a prototype with essential features in order to validate the main idea in a less complex procedure.

\subsection{Setup Procedure}
The developer (website owner) who is intended to enable JSSignature to load a third-party JavaScript, should follow these steps: 
(1) download a copy of JSSignature and upload it to their host, 
(2) use JSSignature to load trusted external scripts by their corresponding valid public keys instead of using script tags in the web pages.

To utilize JSSignature with all advantages, third-party JavaScript provider is required to add the signature to its script. The third-party JavaScript provider follows these steps: 
(1) sign the provided script by appending the generated signature string at the first line of the script as a JavaScript comment, 
(2) and publish the public key of the provided script to be utilized in script inclusion using JSSignature.

The website visitor does not need to do a modification. The end-user will not notice any change in websites those have started using JSSignature.

\subsection{Prototype Implementation}
JSSignature's security strength is based on digital signature principle. Therefore, choosing a right digital signature algorithm and its implementation are essential to achieve the maximum security.

In our prototype, we have chosen SHA-256 algorithm to generate the script digest. SHA-256 algorithm is considered as a safe hashing algorithm to be used in digital signatures at the time of writing this paper. For public-key encryption, RSA algorithm has been chosen.

However, there are no obstacles to choose other algorithms. The mentioned algorithms are the cryptographic foundation of the latest version of Transport Layer Security which is widely-used nowadays on millions of websites as a secure combination \cite{ietf:ssl}. 

\subsubsection{Digital Signatures}
Digital signature is a mathematical scheme to indicate approval of a document. Verification of a signature is possible for anyone, but signing a document is possible just by one identity who has the secret. Moreover, validating the integrity of the document is in the nature of this scheme. If someone changes the signed document, the signature will be no longer valid. Digital signatures are based on public-key cryptography also known as asymmetric cryptography. Digital signature concept was proposed by Diffie and Hellman as equivalent of a written signature \cite{diffie:crypto}. 

A few years next, Rivest, Shamir, and Adleman invented RSA cryptosystem and RSA signature which is widely used nowadays. RSA as a practicable asymmetric cryptosystem uses two distinct keys for encryption and decryption \cite{rivest:publickey}. For signing a large document, it is necessary to produce a short digest and sign it behalf of the original document to achieve efficiency, integrity and compatibility. The signed digest will be attached to the document. Digest of document can be generated using a cryptographic hash algorithm (e.g., MD5, SHA-X, RIPEMD). In signature verification phase, digest of document will be generated again to be used in the verification procedure \cite{rivest:digitals}.

Digital signature scheme benefits from the following properties: (1) \textbf{Integrity}. We can make sure the document has not altered compared to the signed version, (2) \textbf{Authentication}. The identity who has the secret has signed the document. It is possible to validate the source of a document, (3) \textbf{Non-repudiation}. The identity who has signed the document, cannot deny it in the future. It brings responsibility toward script contents \cite{rivest:publickey}.

Majority of cryptography algorithms are already implemented in JavaScript language and are available with free licenses.
In this prototype, CryptoJS \cite{cryptojs} as a custom MIT licensed SHA-256 hashing implementation in JavaScript has been used. For the asymmetric encryption, RSA algorithm library implementation by Tom Wu \cite{rsa} is chosen. There will be no obstacle to choose other implementations of cryptography algorithms to gain a better performance or for any other reason. In this prototype, the selection criteria for the cryptography JavaScript implementations was based on two factors of (1) ease of integration, (2) and widespreadness. \textit{Web Cryptography API} is recently introduced as a web browser feature \cite{w3c:crypto}.  It would significantly improve JSSignature performance, reliability and lower code complexity. However, we have ignored it as it is not still available wide-spread in all major web browsers on the date of writing this paper.

\subsubsection{Fetch the Third-Party JavaScript Resource}
JSSignature works as a client-side proxy in loading third-party JavaScript libraries. Therefore, it has to fetch the remote hosted file in RAW data for further processing and possibly executing it as a JavaScript code in the web browser. The remote file will be downloaded and stored in a JavaScript variable by using AJAX requests as a string. AJAX required functions are working in all modern standard browsers, however, requesting a resource from a different origin is limited by web browsers.

According to the nature of JSSignature, all its requests are from other domains those are providing the JavaScript libraries. To enable client-side cross-origin requests inside web browsers, W3C has introduced Cross-Origin Resource Sharing (\textit{CORS}) \cite{w3c:cors}. At the date of writing this paper, \textit{CORS}  is supported for more than 93\% of global web users \cite{caniuse:cors}. \textit{CORS} must be enabled by the resource provider by adding an HTTP header. \textit{CORS} HTTP headers are currently available and active on hosted JavaScript files in famous JavaScript library CDN providers such as Google Hosted Libraries and CdnJs.com. Anyway, enabling \textit{CORS} fetch for the provided resources is possible for any provider with a straightforward standard procedure in their web server configurations. To satisfy the legacy support requirement, JSSignature is able to fetch \textit{CORS} disabled JavaScript resources by using a server-side proxy. JSSignature will detect \textit{CORS} disabled resources and re-request them through this self-hosted proxy to bypass \textit{SOP} limits.

\subsubsection{Kernel}
JSSignature is coded in pure JavaScript and no other libraries except the mentioned cryptography implementations are included. The whole prototype has been implemented in a single JavaScript file to make the usage simpler.

\begin{lstlisting}
<script src="jsSignature.js"></script>
<script>
 var _publicKey = 'nRx8Ifkiw4hTgFL1Xx...';
 jsSignature.loadJs(_publicKey, 'http://cdn.example.com/jquery.js');
</script>

\end{lstlisting}

JSSignature will fetch the third-party JavaScript resource in the first step. The fetching function requests the resource in the client-side, and in case of failure, it would re-request it through the server-side proxy. To vanish the chance of website failure in case of signature validation problem due to a real threat or technical problem, it is possible to host a local copy of the trusted library as an alternative. JSSignature accepts the alternative JavaScript URL as an optional parameter in loadJs() function. In case of any error, the local script will be inserted into the page via DOM features as a new script tag.

Next, it reads the first line of the fetched content and extracts signature for validation check. The signature is included as a JavaScript comment and it does not impact a legacy inclusion of JavaScript resource.
\begin{lstlisting}
//JSSignature:RgnNFVQ2zsAtnxwbdcUpT508...
/*! jQuery v2.1.1 */
!function(a,b){... //js content
\end{lstlisting}
The extracted signature will be decrypted using the publicly available key. The rest of the content will be hashed and compared with the decrypted signature to validate the inclusion. The code scenario is followed by the procedure discussed in section 3.2.

\section{EVALUATION}
The principle requirements defined in section 2.3 are assessed and evaluated by the implemented prototype. The test reports in this paper are limited to a single third-party JavaScript resource since there is not any technical difference between additional cases due to the nature of JSSignature technique. Table 1 shows the comparison between JSSignature and other state of the art proposed methods to handle the risk.

\subsection{Invulnerability}
JSSignature security measure is linked to digital signature and cryptography, so it has a mathematical foundation. JSSignature validates the source and integrity of the remote JavaScript instead of the script processing. Therefore, any attack to the benign third-party JavaScript will be mitigated, regardless of the attack nature or possible JavaScript vulnerabilities. Therefore, this technique is invulnerable against any kind of threat to the third-party JavaScript resource.

Before validating the JavaScript resource, it loads in the client's machine just as a plain-text and it will be executed only after the signature verification. Consequently, the method cannot be exploited using infected codes or \textit{zero-day} vulnerabilities, as it is not considered as JavaScript codes before validation. To the best of our knowledge, the breakpoint of JSSignature is same as the digital signature breakpoint. 

The suggested technique brings the following digital signature advantages to third-party JavaScript resources: (1) \textit{Integrity}: any unauthorized modification in the JavaScript resource is not possible. So, the attacker is unable to inject or update the code. (2) \textit{Authentication}: the creditable third-party JavaScript provider has the private key and he is the only individual who can sign a modified code. (3) \textit{Non-repudiation}: the creditable third-party JavaScript provider is unable to do a sinister action by himself, as it brings legal responsibility.

\begin{table}
	\caption {A comparision between JSSignature and other state of the art alternative methods. \textit{Compatibility} refers to the compatibility with all available third-party JavaScript libraries and also compatibility with generic web browsers and platforms.} \label{tab:title}
    \begin{tabular}{|p{3.9cm}|p{2.3cm}|p{3.5cm}|p{2.1cm}|}
    \hline
    {Method}                      & {Invulnerability} & {Keeping Third-party JavaScript Advantages} & {Compatibility} \\ \hline
    SRI (Subresource Integrity) & Yes & Partial  & Partial  \\ \hline
    CSP (Iframe Isolation)      & No  & Partial  & Partial  \\ \hline
    COWL                        & No  & Partial  & Partial  \\ \hline
    JSSignature (Our method)                 & Yes & Full & Full \\ \hline
    \end{tabular}
\end{table}
\subsection{Avoiding Losing Third-Party JavaScript Advantages}
JSSignature downloads the third-party JavaScript resource directly from the remote host through the visitor's web browser. All mentioned advantages of remote hosted libraries are satisfied within this method: (1) Web browser cache will be used in case of the existence of the resource in the cache, (2) and as it will be loaded from the remote host, there is no issue regarding the maintenance cost. Due to utilizing digital signatures, the third-party provider is able to update the script regardless of any change in its user's websites. It is enough to re-sign the updated script by using the private key and no further action is required. The new updated resource will be available seamlessly for all the third-party JavaScript resource users. Therefore, the trusted third-party JavaScript provider is able to develop its script without any restriction.

\subsection{Compatibility}
JSSignature is a pure JavaScript library which works inside a webpage without any additional requirement. As a consequence, JSSignature is operation system independent and no additional software or browser add-ons are required and will be fully supported in major web browsers. No modification by the end-user is required.

Utilizing JSSignature to load a validated version of a third-party JavaScript resource is possible for all currently available provided libraries. Achieving to all the discussed technical merits requires the third-party provider to sign the provided resource with JSSignature structure described in section 3.3. However, for legacy support, JSSignature is able to check the third-party resource without a signature and verify it just with the calculated digest of manually reviewed JavaScript resource. JSSignature does not interfere the included JavaScript code after the initial signature validation. Consequently, there will be no compatibility issue or any chance of conflict with available JavaScript libraries. To utilize JSSignature, no code structure modification is required in the third-party JavaScript resources. Moreover, it does not have any side-effect on other users who are not utilizing JSSignature to use the inclusion since the included signature is just a JavaScript comment and it will be ignored by default.

\subsection{Performance}
Performance report is described in two aspects of loading overhead and run-time overhead. In continue, the two aspects will be detailed in the respective subsections. Finally, the performance will be discussed regarding real-world web application requirements.

\subsubsection{Loading Overhead}
For validation purposes, there is a processing time overhead. To load a third-party JavaScript securely through JSSignature, it is required to calculate the SHA-256 of the file, and compare it with the RSA decrypted signature of it. These two cryptography processes are costly and SHA-256 hashing overhead is directly related to the third-party JavaScript resource file size. For performance measuring test case, we have considered the following environment: 
\begin{enumerate}[a.]

\item jQuery as the most famous and a widespread JavaScript library has been chosen as the third-party hosted JavaScript resource. It is provided by the majority of well-known free CDN providers such as Google Hosted Libraries. Moreover, jQuery is one of the heaviest commonly used JavaScript libraries in both aspects of file size and code complexity. The minified 82.2kb version of latest jQuery has been used.

\item Test cases have been executed on a personal computer with Intel Core i7-4702MQ 2.2GHz processor and 16 GiB of DDR3 RAM.

\item Chrome web browser has been used in testing because of the powerful reporting features and being developer friendly.
\end{enumerate}

To get performance statistics, we have implemented an HTML test page to load the jQuery library from a remote host through JSSignature with real-world settings and security checks. The test web page has been reloaded for 1000 times and the loading time has been recorded. To eliminate web browser content request overhead in measurements, the execution time of JSSignature after content fetch has been also measured. The total time, including the browser latency to response the requested file content and executing the validated remote JavaScript library code is 65ms. The execution time overhead directly related to JSSignature process is 30ms, which includes the time required for SHA-256 hashing, extracting the signature from fetched resource, decrypting it by using public key via RSA algorithm and other functions related to the JSSignature validation until the requested JavaScript code execution. Table 2 shows the processing time overheads for all steps.
\begin{table}
	\caption {The processing time overhead for the test case script} \label{tab:title}
\begin{tabular}{|p{11cm}|c|}
\hline 
 \textbf{Factor} & \textbf{Time} \\ 
\hline 
Library Initialize -- including web browser cache reload overhead & 5ms \\ 
\hline 
Verification execution overhead & 30ms \\ 
\hline 
Total overhead & 35ms \\ 
\hline 
\end{tabular} 
\end{table}
\subsubsection{Run-Time Overhead}
JSSignature verifies the third-party JavaScript library before the execution and it does not alter or control the original script after the loading. Therefore, the included JavaScript code works without any compatibility issue or process overhead.

\subsubsection{Performance Discussion}
The most advantage regarding JSSignature performance is the lack of any overhead while script execution. This factor is crucial in modern client-intensive web applications with heavy processes in the client-side. Any execution overhead because of reference monitoring or altering the original script can cause serious negative side-effects. Regarding the initial first-time load process overhead, 35ms total overhead for a jQuery library is satisfying for most usages. This performance is not only considered efficient in compare to the other similar solutions, but it also is negligible for real-world usages. Moreover, this initial overhead can be also optimized by using alternative hashing algorithms or local caching. JSSignature has a minimal overhead in loading the remote hosted resource. It causes no overhead after the initial validation and does not have any performance penalty while executing JavaScript codes.

\section{RELATED WORKS}
Much of the efforts related to this paper are working on handling the possible malware in the included JavaScript resource by using language restrictions, type systems and browser modification. JSSignature presents a different approach based on digital signatures to validate the source and integrity of the remote script instead of handling the injected vulnerability.

\textbf{In-page client-side or server-side script processing}. To protect the web page from the possible threats by a third-party JavaScript resource, several solutions have been suggested by restricting the third-party JavaScript resource to a safe subset of trusted code. Techniques such as ADSafe \cite{adsafe} and ADsafety \cite{adsafety} are validating the untrusted JavaScript codes to be within a safe subset. Some approaches such as such Caja \cite{caja}, GateKeeper \cite{gatekeeper}, Microsoft Web Sandbox \cite{websandbox} and Jacaranda \cite{jacaranda} are re-writing and analyzing the third-party inclusion in the server-side or by a custom developed browser plugin as an alternative \cite{websandbox}. In advanced to transforming the JavaScript codes, some techniques such as BrowserShield \cite{browsershield} are inserting extra checks for possible run-time attacks. 

Transforming a JavaScript code has serious disadvantages: (1) the compatibility issue for complicated JavaScript libraries is a concern. Any change in a JavaScript resource may lead to a failure or a performance issue, and there is no guarantee if the transformed code is working properly with the same efficiency of untouched version. (2) Some of the techniques are requiring server-side processing \cite{caja, websandbox, browsershield, jacaranda}. Therefore, they will ruin the discussed CDN and client-side caching advantage of using a third-party resource. Moreover, any server-side process overhead for a web application is costly, and in a real world web application it is not feasible in most cases. (3) After all, there is still a chance to abuse whitelisted JavaScript codes for a sinister purpose.

The suggested technique in this paper will be executed on the client-side and does not alter the third-party JavaScript. Also, the approach is not about the included third-party JavaScript functions and codes. Instead, the validation of the whole included code block will be verified.

\textbf{Browser modification}. Some techniques require a browser modification to enable protection. Adsentry \cite{adsentry} is implemented as a Firefox add-on, while ConScript \cite{conscript}, WebJail \cite{webjail} and Contego \cite{contego} are implemented on browser JavaScript engine to set policies and inline reference monitoring. MashupOS \cite{mashupos} is a custom web browser which isolates the untrusted inclusions, while allowing whitelisted communications in the browser level. Policies abuse, tight restrictions for benign scripts, and possible vulnerabilities are still the drawbacks.

Content Security Policy (CSP) \cite{csp} as another approach, introduces a new HTTP header to be supported by the web browser as a security feature. Based on this HTTP header, the browser should limit the page resources, including JavaScript files and resources to the whitelisted origins. It is proposed to protect the page from sending private information to an unverified origin, and it does not not prevent policies abuse, and/or other JavaScript vulnerabilities and infections.

An approach based on browser modification has its own advantages and drawbacks. A browser modification may lead to a better efficiency. However, the threat would not be mitigated until all the web application visitors do install the modified web browser which is not practicable. According to the fact that there are various versions of web browsers in the market, releasing a security feature as a standard for all these browsers is unfeasible in an acceptable time period. JSSignature does not require a browser modification. It works with default browser features in the client-side, and the website owner is able to utilize it without any update required for the end-users.

\textbf{Isolating by browser features}. Some approaches are suggested based on browser features for isolating the third-party JavaScript. AdJail \cite{adjail} uses browser features for apply access control. It loads the script and whitelisted elements on a shadow page on another origin to avoid unwanted access based on \textit{SOP}. Any change to an element in the copied version or the original page will be applied to the other one if policies allow it.
Complete isolation is possible by browser standard features. Recent browsers are supporting sandbox attribute for iframe elements to block any access from iframe content. Moreover, \textit{SOP} gives the web developer to load a script in a web page from another origin with full isolating.

JSand \cite{jssand} works in the client-side, and it is based on sandboxing principle. It uses object-capable JavaScript engine in modern web browsers to sandbox and apply the access control architecture. Like all other similar methods, the policies setup and abuse is a major concern.

To conclude, these sandboxing techniques are not fully practicable in real-world use cases for majority of third-party JavaScript libraries. Isolation may make a third-party JavaScript broken, or at least it sets strict limits for a creditable third-party JavaScript provider. A third-party JavaScript is included to work inside the web page and integrate with elements and other JavaScript codes, and isolation is in contrast with this principle. Also, the policies abuse is always a concern. On the other hand, their protection method cannot be confidently verified for all cases and vulnerabilities since their architecture is sensitive to the included JavaScript codes functionality. Therefore, verifying the protection against all available codes and \textit{zero-day} vulnerabilities is unfeasible in similar methods.

\section{CONCLUSION AND FUTURE WORKS}
A prototype of JSSignature has been implemented and has been evaluated. The performance benchmarks were satisfying for real-world usages, and security measures are trustable thanks to the cryptography foundation. This technique is not based on JavaScript code processing, and instead the source and integrity of JavaScript inclusion will be verified.

Consequently, it is invulnerable regardless of the infection nature. Generally, the expected essential requirements of invulnerability, backward compatibility, avoid losing third-party advantages and acceptable performance penalty have been satisfied.
 
To the best of our knowledge, JSSignature is the first technique which protects the web page from any unauthorized third-party JavaScript resource modification or any attack to the resource provider, while it does not set any restriction for the benign provider. However, JSSignature is working based on digital signature scheme and the provider must be creditable to avoid signing a malicious code deliberately without the fear of legal responsibility.

In the future, we are intended to add an option to the prototype's digital signature algorithm to validate signature through Certificate Authorities to vanish public key publishing procedure. Moreover, we would like to transform this technique to a World Wide Web standard to verify all the third-party resources, and not only limited to the JavaScript resources.

\subsubsection*{Acknowledgments}
The research was partially supported by OP RDE project No. CZ.02.2.69/0.0/0.0/16\_027/0008495, International Mobility of Researchers at Charles University.\\
The authors gratefully acknowledge the contribution and helps of Dr. Mehdi Sheikhalishahi and Mahshid Nikravesh to this work.

%
%
%
%

\renewcommand{\bibsection}{\section*{References}}
\bibliographystyle{splncs04}
\begingroup
  \ifluatex
  \else
    \microtypecontext{expansion=sloppy}
  \fi
  \small 
  \bibliography{jssignature}

\begin{thebibliography}{10}
\providecommand{\url}[1]{\texttt{#1}}
\providecommand{\urlprefix}{URL }
\providecommand{\doi}[1]{https://doi.org/#1}

\bibitem{jssand}
Agten, P., Van~Acker, S., Brondsema, Y., Phung, P.H., Desmet, L., Piessens, F.:
  Jsand: Complete client-side sandboxing of third-party javascript without
  browser modifications. In: Proceedings of the 28th Annual Computer Security
  Applications Conference. pp. 1--10. ACSAC '12, ACM, New York, NY, USA (2012).
  \doi{10.1145/2420950.2420952},
  \url{http://doi.acm.org/10.1145/2420950.2420952}

\bibitem{cdnstats}
BuiltWith: Statistics for websites using cdn technologies.
  \url{http://trends.builtwith.com/cdn} (2014),
  \url{http://trends.builtwith.com/cdn}

\bibitem{caniuse:cors}
Caniuse: {Caniuse} cors. \url{http://caniuse.com/cors} (2014),
  \url{http://caniuse.com/cors}

\bibitem{adsafe}
Crockford, D.: Adsafe. \url{http://www.adsafe.org/} (November 2014),
  \url{http://www.adsafe.org/}

\bibitem{diffie:crypto}
Diffie, W., Hellman, M.: New directions in cryptography. IEEE Trans. Inf.
  Theor.  \textbf{22}(6),  644--654 (Sep 2006). \doi{10.1109/TIT.1976.1055638},
  \url{http://dx.doi.org/10.1109/TIT.1976.1055638}

\bibitem{rivest:digitals}
Goldwasser, S., Micali, S., Rivest, R.L.: A digital signature scheme secure
  against adaptive chosen-message attacks. SIAM J. Comput.  \textbf{17}(2),
  281--308 (Apr 1988). \doi{10.1137/0217017},
  \url{http://dx.doi.org/10.1137/0217017}

\bibitem{gatekeeper}
Guarnieri, Salvatore, Livshits, Benjamin: Gatekeeper: Mostly static enforcement
  of security and reliability policies for javascript code. In: Proceedings of
  the 18th Conference on USENIX Security Symposium. pp. 151--168. SSYM'09,
  USENIX Association, Berkeley, CA, USA (2009),
  \url{http://dl.acm.org/citation.cfm?id=1855768.1855778}

\bibitem{jacaranda}
Hopwood, D.S.: Jacaranda. \url{http://jacaranda.org/} (2008),
  \url{http://jacaranda.org/}

\bibitem{mashupos}
Howell, J., Jackson, C., Wang, H.J., Fan, X.: Mashupos: Operating system
  abstractions for client mashups. In: Proceedings of the 11th USENIX Workshop
  on Hot Topics in Operating Systems. pp. 16:1--16:7. HOTOS'07, USENIX
  Association, Berkeley, CA, USA (2007),
  \url{http://dl.acm.org/citation.cfm?id=1361397.1361413}

\bibitem{ietf:ssl}
IETF: Ietf rfc 5246. \url{http://tools.ietf.org/html/rfc5246#section-1.2}
  (August 2008), \url{http://tools.ietf.org/html/rfc5246#section-1.2}

\bibitem{adjail}
Louw, M.T., Ganesh, K.T., Venkatakrishnan, V.N.: Adjail: Practical enforcement
  of confidentiality and integrity policies on web advertisements. In:
  Proceedings of the 19th USENIX Conference on Security. pp. 24--24. USENIX
  Security'10, USENIX Association, Berkeley, CA, USA (2010),
  \url{http://dl.acm.org/citation.cfm?id=1929820.1929852}

\bibitem{contego}
Luo, T., Du, W.: Contego: Capability-based access control for web browsers. In:
  Proceedings of the 4th International Conference on Trust and Trustworthy
  Computing. pp. 231--238. TRUST'11, Springer-Verlag, Berlin, Heidelberg
  (2011), \url{http://dl.acm.org/citation.cfm?id=2022245.2022268}

\bibitem{languageisolation}
Maffeis, S., Taly, A.: Language-based isolation of untrusted javascript. In:
  Proceedings of the 2009 22Nd IEEE Computer Security Foundations Symposium.
  pp. 77--91. CSF '09, IEEE Computer Society, Washington, DC, USA (2009).
  \doi{10.1109/CSF.2009.11}, \url{http://dx.doi.org/10.1109/CSF.2009.11}

\bibitem{conscript}
Meyerovich, L.A., Livshits, B.: Conscript: Specifying and enforcing
  fine-grained security policies for javascript in the browser. In: Proceedings
  of the 2010 IEEE Symposium on Security and Privacy. pp. 481--496. SP '10,
  IEEE Computer Society, Washington, DC, USA (2010). \doi{10.1109/SP.2010.36},
  \url{http://dx.doi.org/10.1109/SP.2010.36}

\bibitem{websandbox}
Microsoft: Microsoft web sandbox. \url{http://www.websandbox.org/} (2010),
  \url{http://www.websandbox.org/}

\bibitem{caja}
Miller, M.S., Samuel, M., Laurie, B., Awad, I., Stay, M.: Caja: Safe active
  content in sanitized javascript (2008),
  \url{http://code.google.com/p/google-caja/}

\bibitem{cryptojs}
Mott, J.: Cryptojs. \url{https://code.google.com/p/crypto-js/} (2015),
  \url{https://code.google.com/p/crypto-js/}

\bibitem{youarewhatyouinclude}
Nikiforakis, N., Invernizzi, L., Kapravelos, A., Van~Acker, S., Joosen, W.,
  Kruegel, C., Piessens, F., Vigna, G.: You are what you include: Large-scale
  evaluation of remote javascript inclusions. In: Proceedings of the 2012 ACM
  Conference on Computer and Communications Security. pp. 736--747. CCS '12,
  ACM, New York, NY, USA (2012). \doi{10.1145/2382196.2382274},
  \url{http://doi.acm.org/10.1145/2382196.2382274}

\bibitem{adsafety}
Politz, J.G., Eliopoulos, S.A., Guha, A., Krishnamurthi, S.: Adsafety:
  Type-based verification of javascript sandboxing. In: Proceedings of the 20th
  USENIX Conference on Security. pp. 12--12. SEC'11, USENIX Association,
  Berkeley, CA, USA (2011),
  \url{http://dl.acm.org/citation.cfm?id=2028067.2028079}

\bibitem{browsershield}
Reis, C., Dunagan, J., Wang, H.J., Dubrovsky, O., Esmeir, S.: Browsershield:
  Vulnerability-driven filtering of dynamic html. ACM Trans. Web  \textbf{1}(3)
  (Sep 2007). \doi{10.1145/1281480.1281481},
  \url{http://doi.acm.org/10.1145/1281480.1281481}

\bibitem{rivest:publickey}
Rivest, R.L., Shamir, A., Adleman, L.: A method for obtaining digital
  signatures and public-key cryptosystems. Commun. ACM  \textbf{21}(2),
  120--126 (Feb 1978). \doi{10.1145/359340.359342},
  \url{http://doi.acm.org/10.1145/359340.359342}

\bibitem{csp}
Stamm, S., Sterne, B., Markham, G.: Reining in the web with content security
  policy. In: Proceedings of the 19th International Conference on World Wide
  Web. pp. 921--930. WWW '10, ACM, New York, NY, USA (2010).
  \doi{10.1145/1772690.1772784},
  \url{http://doi.acm.org/10.1145/1772690.1772784}

\bibitem{webjail}
Van~Acker, S., De~Ryck, P., Desmet, L., Piessens, F., Joosen, W.: Webjail:
  Least-privilege integration of third-party components in web mashups. In:
  Proceedings of the 27th Annual Computer Security Applications Conference. pp.
  307--316. ACSAC '11, ACM, New York, NY, USA (2011).
  \doi{10.1145/2076732.2076775},
  \url{http://doi.acm.org/10.1145/2076732.2076775}

\bibitem{w3c:workers}
W3C: Web workers. \url{http://www.w3.org/TR/workers/} (May 2012),
  \url{http://www.w3.org/TR/workers/}

\bibitem{w3c:cors}
W3C: Cross-origin resource sharing. \url{http://www.w3.org/TR/cors/} (December
  2014), \url{http://www.w3.org/TR/cors/}

\bibitem{w3c:sop}
W3C: Same origin policy.
  \url{http://www.w3.org/Security/wiki/Same_Origin_Policy} (June 2014),
  \url{http://www.w3.org/Security/wiki/Same_Origin_Policy}

\bibitem{w3c:canvas}
W3C: W3c canvas. \url{http://www.w3.org/TR/2014/WD-2dcontext-20140520/} (2014),
  \url{http://www.w3.org/TR/2014/WD-2dcontext-20140520/}

\bibitem{w3c:crypto}
W3C: Web cryptography api. \url{http://www.w3.org/TR/WebCryptoAPI/} (December
  2014), \url{http://www.w3.org/TR/WebCryptoAPI/}

\bibitem{jsuage}
W3Techs: The usage of client-side programming languages for websites.
  \url{http://w3techs.com/technologies/history_overview/client_side_language/all}
  (2014),
  \url{http://w3techs.com/technologies/history_overview/client_side_language/all}

\bibitem{rsa}
Wu, T.: Rsa and ecc in javascript.
  \url{http://www-cs-students.stanford.edu/~tjw/jsbn/} (December 2014),
  \url{http://www-cs-students.stanford.edu/~tjw/jsbn/}

\bibitem{adsentry}
Xinshu, D., Minh, T., Zhenkai, L., Jiang, X.: Adsentry: Comprehensive and
  flexible confinement of javascript-based advertisements. In: Proceedings of
  the 27th Annual Computer Security Applications Conference. pp. 297--306.
  ACSAC '11, ACM, New York, NY, USA (2011). \doi{10.1145/2076732.2076774},
  \url{http://doi.acm.org/10.1145/2076732.2076774}

\end{thebibliography}
\endgroup

\end{document}